\documentclass[journal,10pt]{IEEEtran}
\usepackage[dvips]{graphicx}
\usepackage{amsmath,amsfonts,amssymb}
\usepackage{subeqnarray}
\usepackage{cases}
\usepackage{epsfig}
\usepackage{footmisc}
\usepackage{fixfoot}
\usepackage{stfloats, balance}
\usepackage{algorithm}
\usepackage{algorithmic}
\usepackage{xcolor}

\newtheorem{proposition}{Proposition}

\IEEEoverridecommandlockouts
    \title{Optimal Beamforming for Hybrid Satellite Terrestrial Networks with Nonlinear PA \\ and Imperfect CSIT}
    \author{Chengxiao~Liu,
         Wei~Feng,~\IEEEmembership{Senior Member,~IEEE,}
        Yunfei~Chen,~\IEEEmembership{Senior~Member,~IEEE,}
       Cheng-Xiang~Wang,~\IEEEmembership{Fellow,~IEEE,}
       and Ning Ge, ~\IEEEmembership{Member,~IEEE}
\thanks{
This work was supported in part by the National Natural Science Foundation of
China (Grant No. 61922049, 61771286, 61701457, 91638205, 61960206006), in part by the National Key R\&D Program of China (Grant No. 2018YFA0701601, 2018YFB1801101), in part by the Fundamental
Research Funds for the Central Universities (Grant No. 2242019R30001),
in part by the EU H2020 RISE TESTBED project (Grant No. 734325), in part by the Beijing Natural Science Foundation (Grant No. L172041) and in part by the Beijing Innovation Center for Future Chip.
}
\thanks{C. Liu, W. Feng (corresponding author) and Ning Ge are with the Beijing
National Research Center for Information Science and Technology,
Tsinghua University, Beijing 100084, China. (email: lcx17@mails.tsinghua.edu.cn, fengwei@tsinghua.edu.cn, gening@tsinghua.edu.cn).}
\thanks{C.-X. Wang is with the National Mobile Communications Research Laboratory,
School of Information Science and Engineering, Southeast University,
Nanjing 210096, China, and also with Purple Mountain Laboratories, Nanjing
211111, China. (e-mail: chxwang@seu.edu.cn).}
\thanks{Y. Chen is with the School of Engineering, University of Warwick, Coventry
CV4 7AL, U.K. (e-mail: Yunfei.Chen@warwick.ac.uk).}
}

\allowdisplaybreaks[4]
\begin{document}
    \maketitle
    \begin{abstract}
        In hybrid satellite-terrestrial networks (HSTNs), spectrum sharing is crucial to alleviate the ``spectrum scarcity'' problem. Therein, the transmit beams should be carefully designed to mitigate the inter-satellite-terrestrial interference. Different from previous studies, this work considers the impact of both nonlinear power amplifier (PA) and large-scale channel state information at the transmitter (CSIT) on beamforming. These phenomena are usually inevitable in a practical HSTN. Based on the Saleh model of PA nonlinearity and the large-scale multi-beam satellite channel parameters, we formulate a beamforming optimization problem to maximize the achievable rate of the satellite system while ensuring that the inter-satellite-terrestrial interference is below a given threshold. The optimal amplitude and phase of desired beams are derived in a decoupled manner. Simulation results demonstrate the superiority of the proposed beamforming scheme.
    \end{abstract}

    \section{Introduction}
    \par{
      Nowadays, spectrum sharing in hybrid satellite-terrestrial networks (HSTNs) is attracting more and more research interest. The spectrum sharing technique can not only alleviate the ``spectrum scarcity'' problem, but also provide an opportunity for coordinated system design~\cite{Bankey2018}--\cite{Jia2016}. Under the spectrum sharing regime, inter-satellite-terrestrial interference is inevitable, which usually leads to considerable performance degradation \cite{Baek2016Spectrum,An2016}. Towards this end, beamforming schemes should be tailored for hybrid satellite-terrestrial scenarios~\cite{r1}, rather than only for satellite or terrestrial scenario.
}
\par{
      Khan \emph{et. al.} proposed a semi-adaptive beamforming scheme for HSTNs in~\cite{Khan2012}. Sharma \emph{et. al.} further designed a 3D beamforming method in~\cite{Sharma2015}. The hybrid analog-digital transmit beamforming was further presented in~\cite{Vazquez2018}. These insightful studies have shown the great potential of beamforming for inter-satellite-terrestrial coordination. However, some practical issues were not considered in these works. In practical HSTNs, the nonlinearity of radio-frequency (RF) power amplifiers (PA) and the imperfect channel state information (CSI) are usually inevitable. These non-ideal factors may significantly degrade the performance of HSTNs.
    }
    \par{
       PA nonlinearity often exists in practical systems~\cite{Singya2017Mitigating,saleh1981frequency}, where digital pre-distortion (DPD) modules are widely used to mitigate it~\cite{Layton2017,Beidas2016}. However, both energy consumption and hardware cost are limited for practical HSTNs. Thus, in some cases, PA nonlinearity cannot be fully mitigated. The authors in~\cite{Diaz2005} proposed a joint nonlinear precoding and PA nonlinearity cancellation method for satellite communication systems. In~\cite{Zheng2012}, a beamforming method was re-designed under the generic nonlinear power constraints for satellite-only systems. Due to the coupling interference between satellites and terrestrial systems, these results can not be directly applied to HSTNs.
    }
    \par{
        CSI at the transmitter (CSIT) is another important issue for beamforming design. In~\cite{Khan2012}--\cite{Vazquez2018}, perfect CSIT was assumed. However, the CSIT related to the terrestrial user terminals (UTs) is hard to be perfectly acquired by the satellite in practice. Generally, information exchange between satellites and terrestrial systems requires extra latency and communication resources. Thus, it is difficult to perform channel estimation in an indiscriminate way for both systems~\cite{r8}. This means that it is hard for satellites to acquire the perfect CSI of terrestrial UTs. In contrast, the position-related large-scale CSI can be obtained by satellites in an offline manner with low cost \cite{Feng2017}, which is rather critical in the line-of-sight (LOS) satellite channel environment \cite{Zheng2012}. In our previous work~\cite{Feng2017}, we have used the slowly-varying large-scale CSIT as a typical imperfect CSI condition for resource allocation. Nevertheless, the impact of large-scale CSIT on beamforming remains unknown, to the best of our knowledge.

    }
    \par{
        In this paper, we design a new beamforming scheme for practical HSTNs, considering the impact of both PA nonlinearity and large-scale CSIT. We formulate an
        optimization problem using the Saleh model of PA nonlinearity and the large-scale multi-beam satellite channel parameters. The problem is non-convex and hard to be solved directly. After recasting the original problem by feasible region reduction and variable substitution, an optimal solution to the optimization problem is derived. The performance of the proposed beamforming scheme is then evaluated by simulations. 
    }

       \begin{figure}[t]
    \centering
    \includegraphics[width=3.0in]{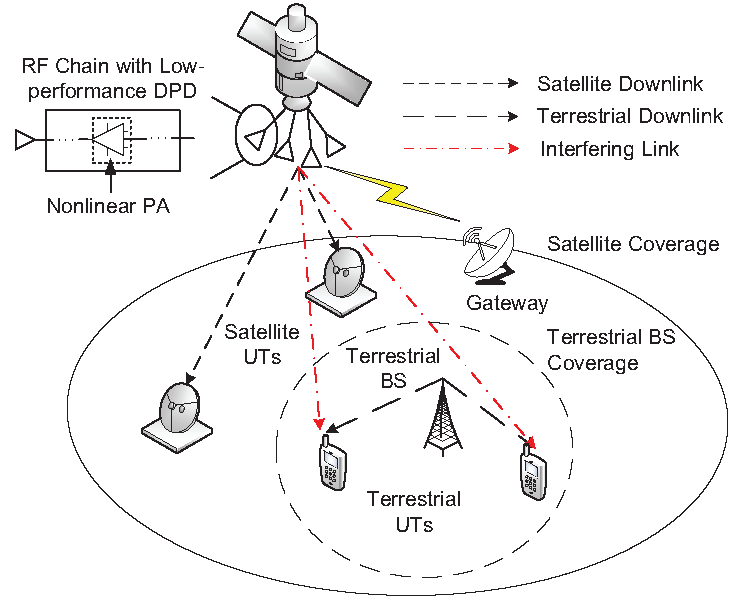}
    \caption{ Illustration of a practical spectrum sharing HSTN. }
    \label{System}
    \end{figure}

    \section{System Model}
    \par{
         We consider a typical spectrum sharing HSTN, as shown in Fig. 1. Regarding the practical issues of HSTNs, an inexpensive nonlinear PA is adopted and imperfect CSIT is assumed. In this case, there are two interfering links \cite{Baek2016Spectrum, Feng2017}. One exists between the satellite and the terrestrial UT, and the other exists between the terrestrial base station (BS) and the satellite UT. Due to the limited coverage area of terrestrial BSs, the latter is usually quite weak, especially when the satellite UT is not covered by terrestrial BSs~\cite{Vazquez2018}. It is worth noting that the latter interfering link is also crucial for some extreme scenarios, e.g. the satellite UTs are close to the terrestrial UTs. In this work, we focus on typical cases, i.e. the satellite and terrestrial UTs are separated to some extent. Hence, we only consider the former interfering link. 

   }

   \par{
        Without loss of generality, we assume that both the satellite UT and the terrestrial UT are equipped with a single antenna for simplicity, and the satellite is equipped with $M$ antennas. After beamforming, all the transmitted signals from these $M$ antennas can be adjusted for better energy efficiency.
        We denote the transmitted symbol as $x=e^{j{\theta}_0}$. Then, with a beamforming vector $\mathbf{w}$,
        the signal vector after beamforming can be expressed as
       \begin{equation} \label{wx}
       \mathbf{\hat{x}} = \mathbf{w} x
       \end{equation}
       where we have $\mathbf{w} = \mathbf{r} \odot e^{j\boldsymbol{\theta}}$, $\mathbf{r} = (r_{1},..,r_{M})^{T}$, $\boldsymbol{\theta}=(\theta_{1},...,\theta_{M})^{T}$ and $\odot$ denotes the Hadamard product of two vectors.


    The signal vector after beamforming, i.e. $\mathbf{\hat{x}}$, will be further amplified via the PA. In practical HSTNs, low-performance DPD modules are adopted to reduce the hardware cost
    \cite{Diaz2005,Zheng2012} so that $\mathbf{\hat{x}}$ is nonlinearly amplified. Particularly, this nonlinearity is modeled by the classic Saleh model~\cite{Singya2017Mitigating,saleh1981frequency}. Such a model can accurately characterize the nonlinear behavior of PAs used for satellite communications \cite{Singya2017Mitigating}.
     Assuming different parameters of the Saleh model for different RF chains, we derive the output signal of PAs as
     \begin{eqnarray}
     &&\!\!\!\!\!\!\!\!\!\!\!\!\!\!\!\!\mathbf{z}(\mathbf{r}, \boldsymbol{\theta}) = \left[z_{1}(r_{1},\theta_{1}),  ...,z_{M}(r_{M},\theta_{M}) \right]^{T} \\
     &&\!\!\!\!\!\!\!\!\!\!\!\!\!\!\!\!z_{i}(r_{i},\theta_{i}) = \frac{\alpha_{i} r_{i}}{1 + \beta_{i} r_{i}^{2}} e^{j \left(\theta_{0} + \theta_{i}  + \frac{\alpha_{\phi_{i}} r_{i}^{2}}{1 + \beta_{\phi_{i}} r_{i}^{2}} \right)},i=1 \sim M \label{saleh}
    \end{eqnarray}
    where $\alpha_{i}, \beta_{i},\alpha_{\phi_{i}}, \beta_{\phi_{i}}$ are parameters of the Saleh model.
}
\par{
       We consider a composite multi-beam satellite channel model, which has been widely used in satellite systems, due to its advantages in characterizing the LOS satellite channel environment  and the correlation among multiple antennas~\cite{Zheng2012}. We denote the channel between the satellite and its UT as $\mathbf{h}_{(s \to s)}$, which can be expressed as
       \begin{equation}
        \mathbf{h}_{(s \to s)} = \sqrt{g_{s}} \xi_{s}^{\frac{1}{2}} e^{-j\phi_{s}}  \mathbf{b}_{s}^{\frac{1}{2}}. \label{hss}
       \end{equation}
       In (\ref{hss}), $g_{s}$ represents the free-space path loss, $\xi_{s}$ denotes the power attenuation of the rain fading, $\phi_{s}$ is a uniformly distributed phase of the antenna feeds, and $\mathbf{b}_{s}$ denotes the beam gain, which physically also contains the correlation among multiple satellite antennas \cite{Zheng2012}. Similarly, we have the interfering link between satellite and the terrestrial UT $\mathbf{h}_{(s \to t)}$ as
         \begin{equation}
        \mathbf{h}_{(s \to t)} = \sqrt{g_{t}} \xi_{t}^{\frac{1}{2}} e^{-j\phi_{t}} \mathbf{b}_{t}^{\frac{1}{2}}. \label{hst}
       \end{equation}
According to \cite{Zheng2012}, $g_{s}$, $g_{t}$, $\mathbf{b}_{s}$, and $\mathbf{b}_{t}$ vary with the location of UT, which remain constant on the order of seconds.
 $\xi_{s}$ and $\xi_{t}$ vary with the atmospheric environment, which remain constant on the order of minutes. In contrast, $\phi_{s}$ and $\phi_{t}$ vary much  faster than the aforementioned parameters. We denote $g_{s}$, $g_{t}$, $\xi_{s}$, $\xi_{t}$, $\mathbf{b}_{s}$, and $\mathbf{b}_{t}$ as large-scale parameters. Then, the large-scale channel gain vector can be derived as
          \begin{eqnarray}
           && \mathbf{l}_{(s \to s)} = \sqrt{g_{s}} \xi_{s}^{\frac{1}{2}} \mathbf{b}_{s}^{\frac{1}{2}} \\
           && \mathbf{l}_{(s \to t)} = \sqrt{g_{t}} \xi_{t}^{\frac{1}{2}} \mathbf{b}_{t}^{\frac{1}{2}}. \label{lst}
        \end{eqnarray}
We denote $\phi_{s}$ and $\phi_{t}$ as small-scale parameters. In practice, we assume that the slowly-varying large-scale CSIT is known for beamforming optimization.}

      \section{Beamforming Optimization}
    \par{
         Based on (\ref{wx})-(\ref{lst}), the received signal at the satellite UT can be expressed as
         \begin{equation}
         y_{s} = \mathbf{h}_{(s \to s)}^{H} \mathbf{z}(\mathbf{r}, \boldsymbol{\theta}) + n
        \end{equation}
        where $n \sim \mathcal{N}(0, \sigma^{2})$ is the additive white Gaussian noise. Then, the achievable rate of the satellite system can be calculated as
        \begin{eqnarray}
            && \!\!\!\!\!\!\!\!\!\!\!\!\!\!\!\! R(\mathbf{r}, \boldsymbol{\theta}) = \log_{2}\left( 1 + \frac{|\mathbf{h}_{(s \to s)}^{H} \mathbf{z}(\mathbf{r}, \boldsymbol{\theta} )|^{2}}{\sigma^{2}} \right) \nonumber \\
         \nonumber  && \ \  = \log_{2}\left( 1 + \frac{\mathbf{z}(\mathbf{r}, \boldsymbol{\theta})^{H} \mathbf{l}_{(s \to s)} e^{-j \phi_{s}} e^{j \phi_{s}} \mathbf{l}_{(s \to s)}^{T} \mathbf{z}(\mathbf{r}, \boldsymbol{\theta})}{\sigma^{2}} \right) \\
           && \ \  = \log_{2}\left( 1 + \frac{|\mathbf{l}_{(s \to s)}^{T} \mathbf{z}(\mathbf{r}, \boldsymbol{\theta} )|^{2}}{\sigma^{2}} \right).
        \end{eqnarray}
        
        To ensure that the inter-satellite-terrestrial interference is below a given threshold $\epsilon$, we have
      \begin{equation} \label{eqn10}
       \mathbf{E}_{\phi_{t}} \left\{ |\mathbf{h}^{H}_{(s \to t)} \mathbf{z}(\mathbf{r}, \boldsymbol{\theta})|^{2} \right\} \leq (\mathbf{l}_{(s \to t)}^{T} \mathbf{\bar{z}}(\mathbf{r}))^2 \leq \epsilon
       \end{equation}
       where $\mathbf{\bar{z}}(\mathbf{r}) = (\frac{\alpha_{1} r_{1}}{1 + \beta_{1} r_{1}^{2}},  ...,\frac{\alpha_{M} r_{M}}{1 + \beta_{M} r_{M}^{2}})^{T}$
       and $\mathbf{E}_{\phi_{t}}$ denotes the expectation with respect to the unknown small-scale channel parameters. The constraint in (\ref{eqn10}) characterizes the upper bound of inter-satellite-terrestrial interference, which has different forms when different channel models are adopted.
    }

    \par {
      We aim to maximize the achievable rate of the satellite system while guaranteeing the inter-satellite-terrestrial interference below a given threshold. The beamforming optimization problem is formulated as follows
    \begin{subequations} \label{prob1}
    \begin{align}
             \max_{\mathbf r, \boldsymbol \theta}  & ~~~\log_{2}\left( 1 + \frac{|\mathbf{l}_{(s \to s)}^{T} \mathbf{z}(\mathbf{r}, \boldsymbol{\theta} )|^{2}}{\sigma^{2}} \right) \label{prob1a} \\
               s.t.    &~~~ (\mathbf{l}_{(s \to t)}^{T}    \mathbf{\bar{z}}(\mathbf{r}))^2 \leq \epsilon \label{prob1b}  \\
                       &~~~ \sum_{i = 1}^{M} r_{i}^{2} \leq P \label{prob1c} \\
                       &~~~ r_{i} \geq 0, i=1 \sim M \label{prob1d}
    \end{align}
    \end{subequations}
     where (\ref{prob1c}) denotes the power constraint of the input signal of PAs. It is easy to prove that this problem is not convex, due to the non-convexity of $\mathbf{z}(\mathbf{r}, \boldsymbol{\theta} )$~\cite{Boyd2004}, so that (\ref{prob1}) is hard to be solved directly. However, the following proposition applies.

 \begin{proposition}
  There exists one optimal solution $(\mathbf{r}^{*}, \boldsymbol{\theta}^{*})$ to the problem in (\ref{prob1}) that satisfies:
  \begin{eqnarray}
  &&\!\!\!\!\!\!\!\!\!\!\!\! \boldsymbol{\theta}^{*} = -\theta_{0} \mathbf{1}_{M} - \left(\frac{\alpha_{\phi_{1}} {r^{*}_{1}}^{2}}{1 + \beta_{\phi_{1}} {r^{*}_{1}}^{2}} ,  ...,\frac{\alpha_{\phi_{M}} {r^{*}_{M}}^{2}}{1 + \beta_{\phi_{M}} {r^{*}_{M}}^{2}}\right)^{T} \label{theta1} \\
    &&\!\!\!\!\!\!\!\!\!\!\!\! r_{i}^{*} \leq \sqrt{\frac{1}{\beta_{i}}}, i=1 \sim M. \label{r1}
\end{eqnarray}
\end{proposition}
\par{
 \!\!\emph{Proof:~}If $(\mathbf{r}^{*}, \boldsymbol{\theta}^{*})$ is an optimal solution to (\ref{prob1}), it is easy to check that $(\mathbf{r}^{*}, \boldsymbol{\theta}^{*} + \phi \mathbf{1}_{M})$ is also an optimal solution for any arbitrarily given $\phi$. Thus, the problem has more than one optimal solutions.  Note that all the constraints in (\ref{prob1}) have no relationship with $\boldsymbol{\theta}$, and (\ref{prob1a}) is maximized with respect to $\boldsymbol{\theta}$ when the phase of all the components of $\mathbf{z}(\mathbf{r}, \boldsymbol{\theta} )$ are aligned. Thus, there must exist one optimal solution that satisfies (\ref{theta1}).

If there is no optimal solution that satisfy (\ref{r1}), taking $(\mathbf{r}^{*}, \boldsymbol{\theta}^{*})$ as an example, there must exist some $1\leq k \leq M$ that satisfies $r^{*}_{k} > \sqrt{\frac{1}{\beta_{k}}}$.
Then, we define
  \begin{eqnarray}
    &&\!\!\!\!\!\!\!\!\!\!\!\! \gamma_{k} = \frac{\alpha_{k} r^{*}_{k}}{1 + \beta_{k}( r^{*}_{k})^{2}} \label{gammak} \\
    &&\!\!\!\!\!\!\!\!\!\!\!\! r^{\star}_{k} = \frac{\alpha_{k} - \sqrt{\alpha^{2}_{k} - 4\beta_{k} \gamma_{k}^{2}}}{2\beta_{k} \gamma_{k}}. \label{rstark}
\end{eqnarray}
It is easy to observe that $r^{\star}_{k} \leq \sqrt{\frac{1}{\beta_{k}}}$.
Replacing all the $r^{*}_{k}$ in $\mathbf{r}^{*}$ with $r^{\star}_{k}$, one may derive another solution $(\mathbf{r}^{\star}, \boldsymbol{\theta}^{\star})$, where $\boldsymbol{\theta}^{\star}$ is updated according to (\ref{theta1}), so that $(\mathbf{r}^{\star}, \boldsymbol{\theta}^{\star})$ surely satisfies all the constraints in (\ref{prob1}). Moreover, from (\ref{gammak}) and (\ref{rstark}), we have
\begin{equation}
   \frac{\alpha_{k} r^{*}_{k}}{1 + \beta_{k} {r^{*}_{k}}^{2}} = \frac{\alpha_{k} r^{\star}_{k}}{1 + \beta_{k} {r^{\star}_{k}}^{2}}.
\end{equation}
Thus, it is easy to find
$R(\mathbf{r}^{\star}, \boldsymbol{\theta}^{\star})=R(\mathbf{r}^{*}, \boldsymbol{\theta}^{*})$,
which indicates that $(\mathbf{r}^{\star}, \boldsymbol{\theta}^{\star})$ is also an optimal solution to (\ref{prob1}). Accordingly, one can conclude that there must exist one optimal solution that satisfies
both (\ref{theta1}) and (\ref{r1}) .
}

 According to \emph{Proposition 1}, the problem in (\ref{prob1}) can be recast without loss of optimality as
    \begin{subequations} \label{prob2}
    \begin{align}
             \max_{\mathbf r, \boldsymbol \theta}  & ~~~\log_{2}\left( 1 + \frac{|\mathbf{l}_{(s \to s)}^{T} \mathbf{\bar{z}}(\mathbf{r})|^{2}}{\sigma^{2}} \right) \label{prob2a} \\
              s.t.    &~~~ (\mathbf{l}_{(s \to t)}^{T} \mathbf{\bar{z}}(\mathbf{r}))^2 \leq \epsilon \label{prob2b} \\
                       &~~~ \sum_{i = 1}^{M} r_{i}^{2} \leq P \label{prob2c} \\
                       &~~~ 0 \leq r_{i} \leq \sqrt{\frac{1}{\beta_{i}}}, i=1 \sim M \label{prob2d} \\
                       &~~~ \theta_{i} = -\theta_{0} - \frac{\alpha_{\phi_{i}} {r_{i}}^{2}}{1 + \beta_{\phi_{i}} {r_{i}}^{2}}, i=1 \sim M.\label{prob2e}
    \end{align}
    \end{subequations}
 Due to the constraints in (\ref{prob2d}) and (\ref{prob2e}), the feasible region of the problem is reduced. However, as \emph{Proposition 1} implies, we can still find an optimal solution. More importantly, it is observed that we
can obtain an optimal amplitude $\mathbf{r}^{*}$ and the corresponding optimal phase $\boldsymbol{\theta}^{*}$ in a decoupled manner, because the variable $\boldsymbol \theta$ only exists in (\ref{prob2e}). Hence, the key challenge is to find $\mathbf{r}^{*}$.

As (\ref{prob2}) is non-convex, it is difficult to derive the optimal solution directly. To handle this problem, let $\mathbf{\bar{z}} = (\bar z_1,\bar z_2,...,\bar z_M)^T$, we give the following optimization problem,
    \begin{subequations} \label{prob3}
    \begin{align}
             \max_{\mathbf{\bar{z}}}  & ~~~\mathbf{l}_{(s \to s)}^{T} \mathbf{\bar{z}} \label{prob3a} \\
              s.t.    &~~~ \mathbf{l}_{(s \to t)}^{T} \mathbf{\bar{z}}\leq \sqrt \epsilon \label{prob3b} \\
                       &~~~ \sum_{i = 1}^{M} \left[\frac{\alpha_{i} - \sqrt{\alpha^{2}_{i} - 4\beta_{i}{\bar z}^{2}_{i}}}{2\beta_{i} \bar z_{i}}\right]^{2} \leq P \label{prob3c}\\
                       &~~~ 0 \leq \bar z_{i} \leq \frac{\alpha_{i}}{2\sqrt{\beta_{i}}}, i=1 \sim M. \label{prob3d}
    \end{align}
    \end{subequations}

Then (\ref{prob2}) can be solved based on the solution to (\ref{prob3}) and the following proposition.

\begin{proposition}
  The problem shown in (\ref{prob3}) is convex. Denoting the optimal solution to (\ref{prob3}) as $\mathbf{\bar{z}}^{*}$, one optimal $\mathbf{r}^{*}$ can be obtained as
     \begin{eqnarray}
r_i^{*} = \frac{\alpha_{i} - \sqrt{\alpha^{2}_{i} - 4\beta_{i}
(\bar{z}^{*}_{i})^{2}}}{2\beta_{i} \bar z_{i}^{*}}, i=1 \sim M. \label{rstari}
    \end{eqnarray}
\end{proposition}
\par{
\emph{Proof:~}It is easy to see that
given (\ref{prob2d}), ${\bar z}_{i}= \frac{\alpha_{i} r_{i}}{1 + \beta_{i} r_{i}^{2}}$ is a monotonically increasing function of $r_{i}$. Performing variable substitution, one can derive (\ref{prob3}) from (\ref{prob2}), as well as the inverse relationship shown in (\ref{rstari}). Hence, (\ref{prob2}) can be solved using the optimal solution to (\ref{prob3}) and the equation in (\ref{rstari}).

Then we prove that (\ref{prob3}) is convex. Define
\begin{equation}
   f(\mathbf{\bar{z}}) = \sum_{i = 1}^{M} \left[\frac{\alpha_{i} - \sqrt{\alpha^{2}_{i} - 4\beta_{i}{\bar z}_{i}^{2}}}{2\beta_{i} \bar z_{i}}\right]^{2} - P
\end{equation}
and
\begin{equation}
   \nu_{i}(x) = \frac{\alpha_{i} - \sqrt{\alpha^{2}_{i} - 4\beta_{i}x^{2}}}{2\beta_{i} x}, i=1 \sim M.
\end{equation}
One further derive
    \begin{eqnarray}
     &&\!\!\!\!\!\!\!\!\!\!\!\!\!\!\!\!\frac{\partial f(\mathbf{\bar{z}})}{{\partial \bar z}_{i}}= \frac{2\nu_{i}(\bar{z}_{i})(1 / \beta_{i} +        \nu_{i}(\bar{z}_{i})^{2})^{2}}{\alpha_{i}(1 / \beta_{i} -  \nu_{i}(\bar{z}_{i})^{2}) } , ~~i=1 \sim M\\
 \nonumber &&\!\!\!\!\!\!\!\!\!\!\!\!\!\!\!\!\frac{\partial^2 f(\mathbf{\bar{z}})}{{\partial \bar z}_{i}^2}= \frac{2 \beta_{i}^{2}(1 / \beta_{i} +  \nu_{i}(\bar{z}_{i})^{2})^{3}}{\alpha_{i}^{2} (1 / \beta_{i} -  \nu_{i}(\bar{z}_{i})^{2})^{3}} \left[ 1 / \beta^{2}_{i} - \nu_{i}(\bar{z}_{i})^{4} \right.\\
&&\!\!\!\!\! \phantom{=\;\;}\left. + 2 \nu_{i}(\bar{z}_{i})^{2}(3 / \beta_{i} -  \nu_{i}(\bar{z}_{i})^{2}) \right], ~~i=1 \sim M\\
     &&\!\!\!\!\!\!\!\!\!\!\!\!\!\!\!\!\frac{\partial^2 f(\mathbf{\bar{z}})}{\partial{\bar z}_{i}{\bar z}_{j}}= 0,~~ i,~j=1 \sim M, i\neq j.
    \end{eqnarray}
Considering (\ref{prob3d}), it is easy to find that $\frac{\partial^2 f(\mathbf{\bar{z}})}{{\partial \bar z}_{i}^2} \geq 0$ for $i=1 \sim M$.
Thus, the Hessian matrix of $f(\mathbf{\bar{z}})$ is a diagonal positive definite matrix. Further considering the obvious convexity of (\ref{prob3a}), (\ref{prob3b}), and (\ref{prob3d}), we see that
(\ref{prob3}) is convex \cite{Boyd2004}.
}

 Based on \emph{Proposition 1} and \emph{Proposition 2}, we can solve (\ref{prob3}) with standard convex optimization tools. Then, we can give the optimal amplitude using the optimal solution to (\ref{prob3}) and (\ref{rstari}). Finally, we can derive the corresponding optimal phase using (\ref{theta1}). 

    \section{Simulation Results and Discussions}
        \par{
        In this section, we present simulation results to demonstrate the superiority of the proposed beamforming scheme. For the parameters of Saleh model in (\ref{saleh}), they are set as $\alpha_{i} = 0.9445 + 0.1 u_{i}$,~$\beta_{i} = 0.5138 + 0.1 v_{i}$,~$\alpha_{\phi_{i}} = 4.0033 + u_{\phi_{i}}$,~$\beta_{\phi_{i}} = 9.1040 + v_{\phi_{i}}$, where $\{u_{i}, v_{i}, u_{\phi_{i}}, v_{\phi_{i}}, i=1,2,...,M \}$ are uniformly distributed over $\left[0, 1\right]$ \cite{saleh1981frequency}. For the satellite channel,  we set $g_{s}=g_{t} = -210 \ \text{dB}$, $\xi_{s}=\xi_{t}=1$, $\phi_{s}$ and $\phi_{t}$ are uniformly distributed in $\left[0, 2\pi\right]$, $\mathbf{b}_{s}$ and $\mathbf{b}_{t}$ are set according to \cite{Zheng2012} with randomly generated locations of UTs. Besides, we set $M  = 16$ and $\sigma^{2}=-107$ dBm.
        }
        \par{
        We compare the proposed beamforming scheme with the conventional maximum ratio transmission (MRT) beamformer and the beamforming algorithm proposed in~\cite{Vazquez2018}. It is worth noting that perfect CSIT was used by the MRT beamformer and the beamformer in~\cite{Vazquez2018}. In the simulation, the MRT beamformer was scaled by a constant to satisfy the interference constraint. Moreover, to the best of our knowledge, PA nonlinearity based on the Saleh model has not been considered in other beamformers, so that it is hard for us to compare the proposed beamformer with other beamformers that ignored PA nonlinearity.
        }

\begin{figure}[t]
\centering
\includegraphics[width=3.2in]{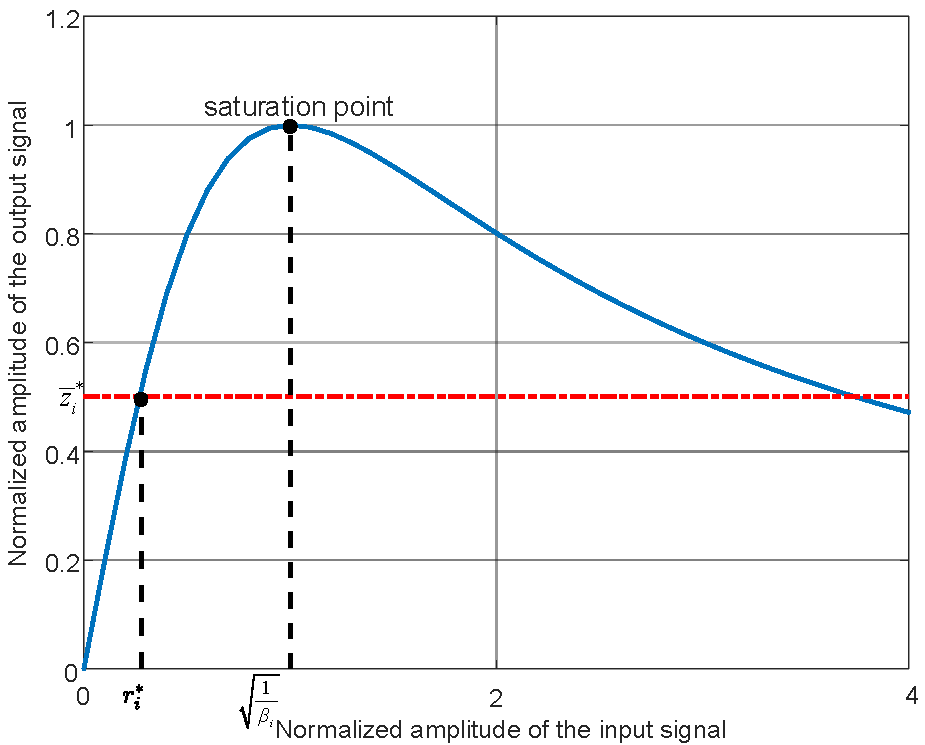}
\caption{A snapshot of the Saleh model that used in the simulation.}
\end{figure}
\begin{figure}[t]
\centering
\includegraphics[width=3.2in]{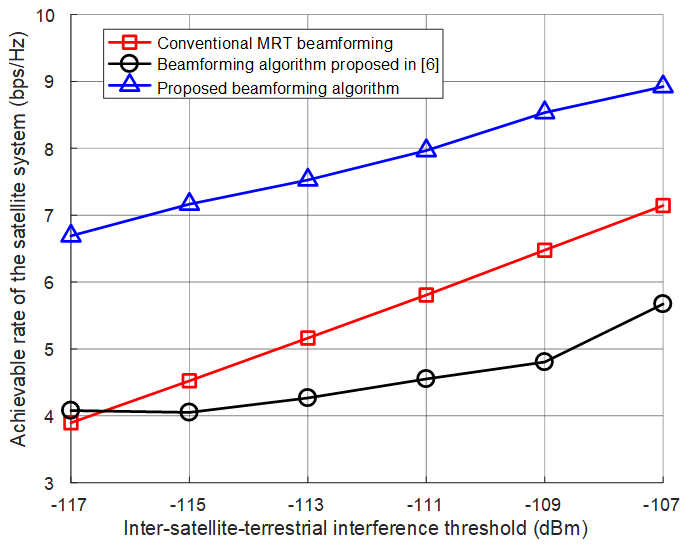}
\caption{Achievable rate of the satellite system with different beamforming schemes when the input power limit $P = 12$ dBw.}
\end{figure}
\begin{figure}[t]
\centering
\includegraphics[width=3.2in]{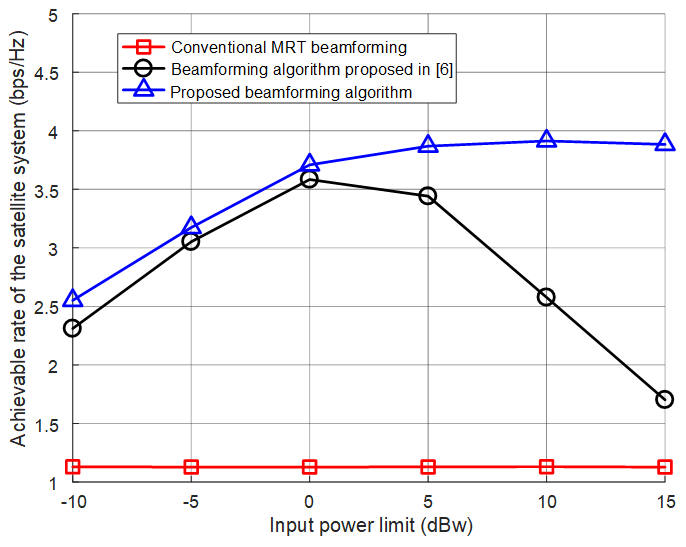}
\caption{Achievable rate of the satellite system with different beamforming schemes when the interference threshold $\epsilon = -107$ dBm.}
\end{figure}
\begin{figure}[t]
\centering
\includegraphics[width=3.2in]{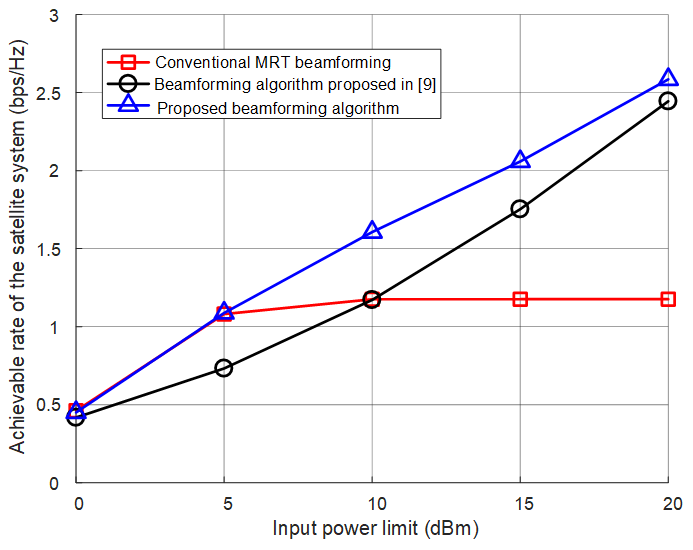}
\caption{Achievable rate of the satellite system with different beamforming schemes when the input power is extremely low with $\epsilon = -107$ dBm.}
\end{figure}
\par{
    In Fig. 2. we discuss the properties of PA nonlinearity and the proposed beamforming optimization scheme. We give a snapshot of the Saleh model used in this simulation. Recalling that the key point of the proposed scheme is to find the optimal amplitude of the beamforming vector, we concentrate on the nonlinearity of amplitude in Saleh model. As shown by the curves, the PA is saturated when $r_{i} = \sqrt{\frac{1}{\beta_{i}}}$. Below the saturation point, we can find the optimal solutions to (\ref{prob2}) and (\ref{prob3}), which satisfy the relationship in (\ref{rstari}). When the amplitude of the input signal increases over the saturation point, the output power of PA decreases correspondingly.
}

     \par{
      In Fig. 3, we consider the achievable rate of the satellite system with different beamforming schemes when the input power limit equals to 12 dBw. One sees that, with the increase of the inter-satellite-terrestrial interference threshold, a larger achievable rate is obtained. Besides, the proposed scheme always outperforms other beamformers, because the proposed scheme jointly considers the PA nonlinearity and large-scale CSIT. Furthermore, the beamformer in \cite{Vazquez2018} has the worst performance. The reason is that the interference constraint was not appropriately considered by the beamformer in \cite{Vazquez2018} under the influence of PA nonlinearity.
      }
      \par{
      In Fig. 4, we evaluate the performance of different beamforming schemes varying with the input power limit of PAs, where $\epsilon = -107$ dBm. As shown by the curves, the interference constraint actually dominates the performance of the MRT beamformer. One can further observe that when the input power limit is lower than 0 dBw, the proposed algorithm provides performances similar to the beamformer in \cite{Vazquez2018}. The reason is that the effect of PA nonlinearity is not significant when the input power is low. Moreover, the advantage of the proposed algorithm in this region comes from the better adaptation to the large-scale CSIT. When the input power increases, the performance gap grows larger. The reason is that the scheme proposed in \cite{Vazquez2018} tends to focus the power on the antennas with larger channel gains. However, recalling the curves in Fig. 2, these focused power will exceed the saturation points of PAs when the input power is high. In this case, a significant reduction in the output signal power of PAs is caused, which can induce the severe performance degradation.
     }
     \par{
        In Fig. 5, we further discuss the performance of different beamforming schemes with varying input power limits of PAs when the input power is extremely low, where $\epsilon = -107$ dBm. We can observe that when the input power is higher than $5$ dBm, the interference constraint still dominates the performance of MRT beamformer, similar to Fig. 4. On the other hand, when the input power is lower than $5$ dBm, the power constraint becomes more important, and the performance of conventional MRT beamformer begins to change with the power limit. Moreover, we can see that the proposed beamformer has a similar performance to the MRT beamformer. The reason is that both the PA nonlinearity and the influence of large-scale CSIT are not important when the input power is extremely low. Besides, we can further observe that the beamformer in \cite{Vazquez2018} still has the worst performance. This fact shows that PA nonlinearity has a more significant influence on the interference constraint than on the achievable rate.
    }

     \section{Conclusion}
    \par {
In this paper, we have investigated the optimal beamforming design with PA nonlinearity and large-scale CSIT for a practical spectrum sharing HSTN. The formulated problem is non-convex. We have solved it using feasible region reduction and variable substitution techniques, and the optimal amplitude and phase of satellite beams have been derived in a decoupled manner. Simulation results have shown that it is valuable to redesign the beamformers to accommodate practical constraints.
}

\bibliographystyle{ieeetr}

\end{document}